\documentclass{appolb}

\newcommand{\be}{\begin{equation}}
\newcommand{\ee}{\end{equation}}
\newcommand{\beeq}{\begin{eqnarray}}
\newcommand{\eeeq}{\end{eqnarray}}

\usepackage{epsfig}
\begin{document}
\title{DGLAP evolution in the saturation model%
\thanks{Presented by K. Golec--Biernat at the X International Workshop on
        Deep Inelastic Scattering (DIS2002), Cracow,
        30 April - 4 May 2002. \\
        Supported by the Polish KBN grant No. 5 P03B 144 20 and by
        the  Deutsche Forschungsgemeinschaft.}
}
\author{J. Bartels${}^1$, K. Golec-Biernat${}^{1,2}$ and H. Kowalski${}^3$
\address{${}^1$II Institut f\"ur Theoretische Physik, Universit\"at Hamburg, Germany\\
${}^2$Instytut Fizyki J\c{a}drowej im. H. Niewodnicza\'nskiego, Krak\'ow, Poland\\
${}^3$Deutsches Electronen Synchrotron DESY, Hamburg, Germany}
}
\maketitle
\begin{abstract}
A modification of the saturation model of deep inelastic scattering at small $x$
which includes the Altarelli-Parisi (DGLAP) evolution is presented.
Significant improvement of the description of the structure function $F_2$
at large $Q^2$ is achieved and a good description of diffractive data is preserved.
\end{abstract}

\section{Introduction}
The saturation model \cite{GBW1} has provided a successful
description of HERA deep inelastic scattering data.
This includes both the $F_2$ data,
with the transition into the nonperturbative
photoproduction region, and  the DIS diffractive data \cite{GBW2}.
Whereas the formulae
are particularly appealing through their simplicity, they also have
an attractive theoretical background: the idea of parton saturation.
Despite its success, the model suffers from the lack of proper scaling violation,
i.e. at larger values of $Q^2$ it does not exactly match with QCD evolution
described by the DGLAP equations.
This becomes clearly visible in the region $Q^2>20$ GeV$^2$ where the
model predictions are below the data on $F_2$. Therefore, one expects that
QCD evolution
should enhance the cross section in this region.

We present a modification of the saturation model which preserves
its success in the low-$Q^2$ (transition) region,
while incorporating QCD evolution in the large-$Q^2$ domain \cite{BGK}.
Since the energy dependence in the large-$Q^2$
region is mainly due to the behaviour of the dipole cross section at small
dipole sizes $r$, our changes will affect mostly the small-$r$ region.
At the same time, we leave the  large-$r$ behaviour of the dipole cross section
practically  unchanged which  allows to retain a good description of DIS diffractive
cross section.
Recent attempts
\cite{LL} along the same lines indicate that diffraction
provides a highly nontrivial restriction on possible modifications of the
saturation model.

%%%%%%%%%%%%%%%%%%%%%%%%%%%%%%%%%%%%%%%%%%%%%%%%%%%%%%%%%%%%%%%%%%%%%%%%%%%%%%%%%%
\section{The saturation model and its modification}

We start with a
brief review of the saturation model
\cite{GBW1}.
Within the dipole formulation of the $\gamma^*p$ scattering, the cross section
\be
\label{eq:totalcross}
\sigma_{T,L}^{\gamma^*p}(x,Q^2)\,=\,
\int d^2r\, dz\;\; \psi^*_{T,L}(Q,r,z)\; \hat\sigma (x,r)\;
\psi_{T,L}(Q,r,z),
\ee
where $\psi_{T,L}$ are  the virtual photon wave functions
with the transverse and longitudinal polarisation. In the saturation model
the following form of the dipole cross section $\hat\sigma$ is proposed
\be
\label{eq:sighat}
\hat\sigma (x,r)\,=\,\sigma_0\,\left\{
1\,-\,\exp\left(-{r^2}/{4 R_0^2(x)}\right)
\right\}\,,
\ee
where $R_0(x)$ is the saturation scale which decreases when $x\rightarrow 0$,
\be
\label{eq:satscale}
R_0^2(x)\,=\,\frac{1}{\mbox{\rm GeV}^2}\,
\left(\frac{x}{x_0}\right)^\lambda\,.
\ee
In order to be able to study the formal photoproduction limit,
the Bjorken variable $x=x_{B}$ was modified to be
\be
\label{eq:bjv}
x\,=\,x_{B}\left(1+\frac{4m_q^2}{Q^2}\right)\,=\,\frac{Q^2+4m_q^2}{W^2}\,,
\ee
where $m_q$ is an effective quark mass
and $W$ denotes the $\gamma^* p$
center-of-mass energy.
The parameters  of the model,
$\sigma_0=23~{\rm mb}$, $\lambda=0.29$ and
$x_0=3\cdot 10^{-4}$ (for fixed
$m_q=140~\mbox{\rm{MeV}}$), were found from a fit to small-$x$ data \cite{GBW1}.

As it is well known \cite{DIPGLUE}, in the small-$r$ region
the dipole cross section
is related to the gluon density obeying the DGLAP evolution
\be
\label{eq:smallr}
\hat\sigma(x,r) \,\simeq\,
\frac{\pi^2}{3}\,r^2\,\alpha_s\,xg(x,\mu^2)\,,
\ee
where the scale $\mu^2\simeq C/r^2$ for $r\rightarrow 0$. The equation (5)
is valid in the double logarithmic approximation in which the constant $C$
is not determined. Expanding the exponent in eq.~(\ref{eq:sighat}) for
$r\ll 2 R_0(x)$, we find the gluon density in the saturation model
\be
xg(x,\mu^2)\,=\,\frac{3}{4\pi^2\alpha_s}\,\frac{\sigma_0}{ R_0^2(x)}\,.
\ee
For fixed $\alpha_s$ this gluon density is clearly scale independent.
%which  calls for a modification taking into account the DGLAP evolution.
Thus, we have to modify the small-$r$ behaviour of the dipole
cross section to include the DGLAP evolution
and, at the same  time, keep the large-$r$ behaviour unchanged.
This will preserve the idea of saturation,
which reflects unitarity, and allows a good description
of  the diffractive cross section.

Therefore, we propose the following modification of the model (\ref{eq:sighat})
\be
\label{eq:sighatnew}
\hat\sigma (x,r)\,=\,\sigma_0\,\left\{
1\,-\,\exp\left(-\frac{\pi^2\,r^2\,\alpha_s(\mu^2)\,xg(x,\mu^2)}
{3\,\sigma_0}\right) \right\}\,,
\ee
where the scale $\mu^2$ is assumed to have the form
\be
\label{eq:scale}
\mu^{2}\,=\,{C}/{r^2}\,+\,\mu_0^2\,.
\ee
The parameters $C$ and $\mu_0^2$ will be determined from a fit
to DIS data. The scale  $\mu_0\gg \Lambda$ allows to freeze the value of
the gluon distribution for large $r$ at a perturbative scale which
leads to  $\hat\sigma(x,r)\,\approx\,\sigma_0$,
as in the original model.  The transition  between small and large $r$
depends on $x$  but in detail it might  be different from the original formulation.
Thus, the modified form mimics in a consistent way the saturated behaviour
of the dipole cross section.

In a first approximation,
$g(x,\mu^2)$ is evolved with the leading order DGLAP
equation in which  quarks are neglected
in the spirit of the small-$x$ limit. The starting gluon
distribution at the initial scale $Q_0^2=1~\mbox{\rm GeV}^2$
\be
\label{eq:gluon}
xg(x,Q_0^2)\,=\,A_g\,x^{-\lambda_g}\,(1-x)^{5.6}\,,
\ee
where $A_g$ and $\lambda_g$ are another fit parameters.
The exponent  determining the
large $x$ behaviour is motivated by the recent
MRST parameterisation \cite{MRST01}.

For small $r$, the exponential in  (\ref{eq:sighatnew})
can be expanded in powers  of its argument, and  relation
(\ref{eq:smallr}) with the running $\alpha_s=\alpha_s(\mu^2)$ is found.
In contrast to the model (\ref{eq:sighat}),
the rise in $1/x$ now has become $r$-dependent. When inserting $\hat{\sigma}$
into (\ref{eq:totalcross}) and convoluting with the
photon wave function, the integrand peaks near $r \sim 2/Q$ for large
$Q^2$, and the argument
of the gluon density turns into $\mu^2 \sim Q^2$. Consequently,
with increasing $Q^2$, DGLAP evolution will strengthen the rise in $1/x$,
whereas in the original saturation model
the power of $1/x$ had been constant

%%%%%%%%%%%%%%%%%%%%%%%%%%%%%%%%%%%%%%%%%%%%%%%%%%%%%%%%%%%%%%%%%%%%%%%%%%%%%%%%%%%
\section{Fit results and comparison with data}

We performed a global fit to the DIS data with $x<0.01$ in the range
$0.1<Q^2<500~\mbox{\rm GeV}^2$. For the HERA
experiments the new 1996-97 data from H1 and ZEUS  were used~\cite{HERA}
together with  the E665 experiment data \cite{E665} (in total 330 points)
The statistical and systematic errors were added in quadrature.
For a full discussion of fit details see \cite{BGK}. With the fixed
quark mass $m_q=140~\mbox{\rm MeV}$, the value of $\chi^2/N_{df}=1.18$ was
found (for the original model refitted to the new
data  $\chi^2/N_{df} \simeq 3$) with the following fit parameters:
$C=0.26$, $\mu_0^2=0.52~\mbox{\rm GeV}^2$, $A_g=1.2$ and $\lambda_g=0.28$.
In addition, for the agreement with the diffractive
data we fix  the normalization of the dipole cross section  to the original
saturation model value $\sigma_0=23~\mbox{\rm mb}$.

The form of the new dipole cross section (\ref{eq:sighatnew})
is shown in Figure~1 (solid lines)
for different values of $x$. As expected the main modification
in comparison to the model (\ref{eq:sighat}) (dashed lines)
lies in the small-$r$ region. In Figure~2 the global characteristic
of the data description is shown. Namely, we plot the effective slope
$\lambda(Q^2)$, obtained from the  parameterisation
of $F_2$ at small $x$,  $F_2\sim x^{-\lambda(Q^2)}$, for fixed $Q^2$.
For the shown curves, $\lambda$  was computed using the relation
$F_2=Q^2/(4\pi^2\alpha_{em})\,\sigma^{\gamma^*p}_{T+L}$ with the two discussed
forms of dipole cross sections. As expected,   the inclusion of the DGLAP evolution
for small $r$ significantly improves agreement with the data at large $Q^2$ while
at small $Q^2$ the results are practically the same.

\begin{figure}[t]
  \vspace*{-1.0cm}
     \centerline{
         \epsfig{figure=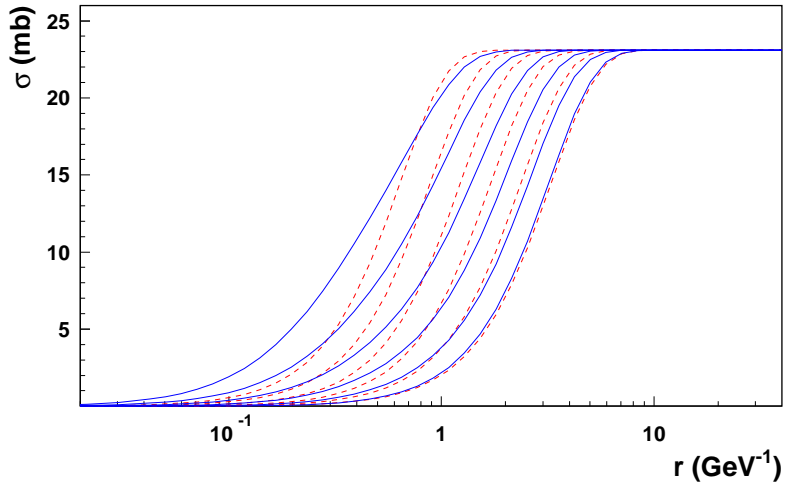,width=12cm}
           }
\vspace*{-0.2cm}
\caption{\it The dipole cross section as a function of the dipole size
$r$ for different
values of $\log_{10} x= -2, -3, ..., -7$ (from the left to the right).
The  original saturation model: dashed lines and the improved model:
solid lines.
\label{fig:1}}
\end{figure}

An important aspect of the dipole models is their straightforward
description of diffractive processes. In particular, the constant
ratio of the inclusive over diffractive cross sections as a function of $x$
finds a natural explanation in the saturation model \cite{GBW1,GBW2}.
In DIS diffraction the cross section is dominated by the contribution
from large dipole sizes $r$. Since the large-$r$  part of the dipole cross
section is practically unchanged in our modification, the description
of diffractive data is as good as in the original saturation model.
The only change introduced is a different treatment of the colour factors
for the $qqg$ component, see \cite{BGK} for more details.

\begin{figure}[t]
  \vspace*{-1.0cm}
     \centerline{
         \epsfig{figure=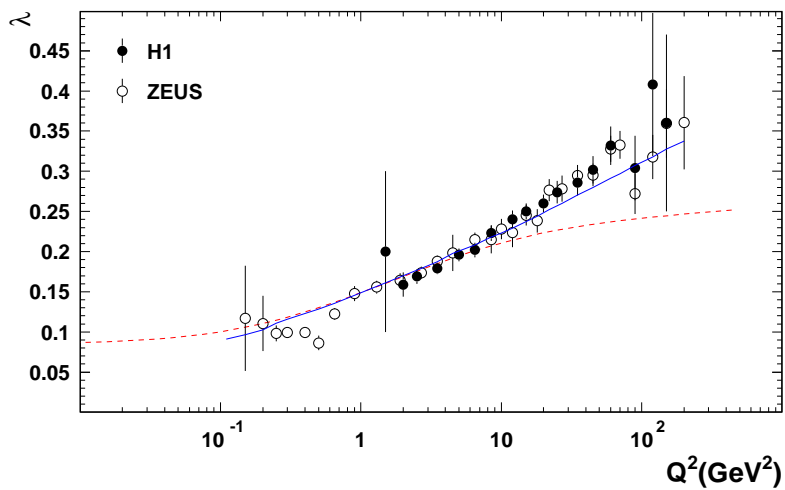,width=12cm}
           }
\vspace*{-0.2cm}
\caption{\it The effective slope $\lambda$
as a function of $Q^2$.
The  original saturation model (dashed line) and the improved model
(solid line) are shown against the data.
%The data are from   ZEUS analysis (open circles)
%and H1 analysis (full circles).
\label{fig:2}}
\end{figure}

\end{document}